\documentclass[12pt]{article} 

\usepackage[margin=1in]{geometry}
\usepackage{amsmath, amsfonts, bm, amssymb}
\usepackage{enumitem}
\usepackage{physics}
\usepackage{subfig}
\usepackage{authblk}
\usepackage[round]{natbib}
\usepackage{graphicx}

\newtheorem{theorem}{Theorem}
\newtheorem{lemma}{Lemma}

\def\y{\bm{y}}
\def\x{\bm{x}}

\def\A{\bm{A}}
\def\W{\bm{W}}
\def\S{\bm{\Sigma}}
\def\g{\gamma}
\def\c{\chi}
\def\ak{\A^{\otimes}}
\def\R{\mathbb{R}}

\begin{document}

\title{A Density Evolution framework for Preferential Recovery of Covariance and Causal Graphs from Compressed Measurements}

\author[1]{Muralikrishnna G. Sethuraman}
\author[2]{Hang Zhang}
\author[1]{Faramarz Fekri}
\affil[1]{School of Electrical and Computer Engineering, Georgia Tech}
\affil[2]{Baidu Research}
\date{}

\maketitle

\begin{abstract}%
In this paper, we propose a general framework for designing sensing matrix $\boldsymbol{A} \in \mathbb{R}^{d\times p}$, for estimation of sparse covariance matrix from compressed measurements of the form $\boldsymbol{y} = \boldsymbol{A}\boldsymbol{x} + \boldsymbol{n}$, where $\boldsymbol{y}, \boldsymbol{n} \in \mathbb{R}^d$, and $\boldsymbol{x} \in \mathbb{R}^p$. By viewing covariance recovery as inference over factor graphs via message passing algorithm, ideas from coding theory, such as \textit{Density Evolution} (DE), are leveraged to construct a framework for the design of the sensing matrix. The proposed framework can handle both (1) regular sensing, i.e., equal importance is given to all entries of the covariance, and (2) preferential sensing, i.e., higher importance is given to a part of the covariance matrix. Through experiments, we show that the sensing matrix designed via density evolution can match the state-of-the-art for covariance recovery in the regular sensing paradigm and attain improved performance in the preferential sensing regime. Additionally, we study the feasibility of causal graph structure recovery using the estimated covariance matrix obtained from the compressed measurements.
\end{abstract}

\section{Introduction}

In this work, we study the feasibility of recovering the covariance matrix and the underlying causal structure of unknown set of variables $\x = (x_1, \ldots, x_p)$, by collecting observations through a linear measurement system of the form,
\begin{equation}
    \y = \A\x + \bm{n},
\end{equation}
where $\y\in\R^{d}$ is of a lower dimension than $\x \in \R^p$. The causal semantics of $\x$ can be represented using a graph $G = (V, E)$ where $V = \{x_1, \ldots, x_p\}$ and the edges encode the dependencies between the variables. The problem recovering the causal structure is then equivalent to the graph structure recovery, in other words, recovery of the edge set $E$. 

Graph structure recovery has been a problem of interest in
the last few decades within the machine learning community.
It is well known that structure recovery is an NP-hard
problem, \citep{np-hard}, and in general it cannot
be uniquely identified \citep{survey}. Nevertheless,
attempts have been made to recover the structure of the
graphical model under various assumptions on the underlying
probability distribution governing the system. Additive
Noise Models (ANM) have gained a lot of traction in recent years due to their analytic simplicity and has been shown
that in such a case the graph structure can be uniquely identified.
In particular, \citet{ligbn} showed
that when the additive noise is Gaussian, the structure of
the Gaussian Bayesian Network (GBN) can be recovered in
polynomial time.

However, the aforementioned solutions assume direct access to
the observational data which may not be practical in certain
applications \citep{comp_moti}, making it an expensive
task to recover the structure of the underlying graph, especially
in high dimensions. Our work differs from the existing methods by considering the scenario where the graph structure is recovered from compressed measurements $\y$ instead of directly observing $\x$. The crux of our approach relies on density evolution analysis
of the message-passing algorithm, also known as Belief
propagation, min-product, or max-sum. The algorithm was
independently developed in different fields in the last century.
In 1935, Bethe \citep{bethe} used it to
approximately compute the partition function. \citet{pearl} developed
belief propagation in 1988 to perform exact inference
in Acyclic Bayesian Networks. 

\subsection{Related Work}. 

\textbf{Compressed Covariance Recovery}. Sparse vector recovery from compressed measurements has been studied quite extensively with several sensing matrices being proposed in the literature \citep{cs1, cs2, nogauss}, each offering some advantages over the others. Over time, Gaussian sensing matrices have become a popular choice for sparse vector recovery. However, \citet{nogauss} showed that the gaussian sensing matrix is not a very good candidate for the recovery of sparse matrices. \citet{binary_vec} showed the use of binary matrices, in particular, adjacency of $\delta$-left regular bipartite graph for sparse vector recovery. \citet{mksetch} built upon the work done by \citet{binary_vec} and proved that adjacency of $\delta$-left regular bipartite graphs can be used for recovery of sparse matrices. 

In the 1960s, \citet{ldpc} proposed a sum-product algorithm to decode \textit{low-density parity check} (LDPC) codes over graphs, which was forgotten for decades and later reinvented along with density evolution to design LDPC codes achieving channel capacity. \citet{statpcs1, statpcs2, statpcs3} analyzed sparse sensing matrices based on spatial coupling using DE for sparse vector recovery. \citet{hang_de} employed density evolution and developed a framework for designing sensing matrices for regular as well as preferential recovery of sparse vectors. For a better understanding of the usage of message passing and density evolution for signal recovery, we refer the readers to \citet{bethe, statpcs3}.  

\textbf{Graph Structure Recovery}. Structure recovery methods for directed graphs can broadly be divided into two categories: (1) Independence test-based, and (2) score-based methods. Independence test-based methods \citep{pcalgo} typically involves computing the conditional independence between any two nodes in the graph conditioned on all the subsets of the remaining nodes. These methods are computationally intensive as the total number of independence tests to be performed grows exponentially in the size of the graph. Moreover, these methods are only capable of finding the graph structure up to Markov equivalency. 

On the other hand, score-based methods rely on a metric to score the candidate \textit{directed acyclic graph} (DAG) based on how well it explains the data. Popular examples of scores are \textit{Akaike Information Criterion} (AIC), Bayesian Information Criterion (BIC) and $\ell_0$ penalized log-likelihood score by \citet{score}. A combinatorial search over the entire space of DAGs is still expensive as the size grows exponentially with the number of nodes in the graph. \citet{notears} proposed a continuous constraint to restrict the search space to that of DAGs using the weighted adjacency matrix and showed its effectiveness for the case of ANM, but their overall optimization program is non-convex and hence not easy to analyze. \cite{ligbn} showed that for ANMs that are also GBNs, the graph structure can be recovered in polynomial time. These are only a few examples from the vast literature available for structure learning, for more information we refer the interested reader to \citet{survey}. 

\subsection{Contributions} This work is the first application of density evolution and message-passing algorithms to design sensing matrices for covariance and graph structure recovery. In particular, we focus on the setting where the covariance is sparse and the parent-child relations are linear. We summarize our contributions as follows:
\begin{enumerate}
\item We propose a novel approach to optimally design a low dimensional data collection (measurement) scheme from a high dimensional signal that would allow for recovering a sparse covariance matrix from these measurements. We use density evolution-based analysis of the message-passing algorithm to reduce the design procedure into a convex program.  
\item The propose two separate design schemes: (i) (\textit{Regular sensing}) equal preference over all the entries of the covariance matrix, and (ii) (\textit{Preferential sensing}) preferential treatment over certain entries of the covariance matrix. We also showcase the feasibility of causal graph recovery from the estimated covariance matrix. 
\item The performance of the proposed sensing systems is validated through numerical simulations.
\end{enumerate}

\textbf{Organization}. In section \ref{sec:Prob-desc} we provide a brief description of the compressed recovery problem, followed by a discussion of the steps involved in designing the sensing system for covariance recovery for regular sensing in section \ref{sec:reg-sensing}, and preferential sensing regime in section \ref{sec:pref-req}. In section \ref{sec:graph-struct-rec}, we discuss the recovery of the graph structure using the estimated covariance matrix. We showcase the effectiveness of our approach via numerical simulations in section \ref{sec:experiments} and end with conclusions in section \ref{sec:conclusion}. 

\section{Problem Description}
\label{sec:Prob-desc}

In this section, we provide a formal description of our problem starting with the notations. All vectors are denoted by lowercase boldface letters, $\bm{x}$, and matrices by uppercase boldfaced letters, $\bm{A}$. $\A_{*, i}$ denotes the $i$-th column of the matrix $\A$, similarly $\A_{j, *}$ denotes the $j$-th row of $\A$. $\norm{\A}_1 = \sum_{ij} |A_{ij}|$ and $\norm{\A}_F = \sqrt{\sum_{ij} A_{ij}^2}$ and $\norm{\x}_p = \big(\sum_{i} x_i^p\big)^{1/p}$.   

Consider a linear measurements system of the form, 
\begin{equation}
\bm{y} = \bm{Ax} + \bm{n},
\label{eq:lin-meas}
\end{equation}
where $\bm{y} \in R^{d}$ denotes the observations, $\x \in R^p$ denotes the unknown vector, $\A \in R^{d \times p}$ denotes sensing matrix, and $\bm{n} \in R^d$ denotes the measurement noise. For the case when $d < p$, we are interested in the problem of recovering the covariance of $\x$ from the observations $\y$. Our goal is to design a sparse sensing matrix $\A$ that is capable of recovering the covariance from compressed measurements and at the same time being able to provide selective preference to a sub-block of the covariance matrix. That is, we would like a sub-block of the covariance matrix to be recovered with a lower probability of error than the rest of the covariance.

Additionally, we model the unknown signal using a \textit{Structural Equation Model} (SEM) \citep{sem1, sem2} given by,
\begin{equation}
x_i = \bm{W}_{*,i}^T\bm{x} + z_i, \quad\forall i = 1, \ldots, p
\label{eq:sem}
\end{equation} 
where $\bm{W}$ denotes the weighted adjacency matrix and $z_i$ corresponds to intrinsic noise in the system. We would also like to learn the weighted adjacency matrix from the compressed measurements $\y$ using the recovered covariance of $\x$. An equivalent representation of the above SEM is to consider a directed (causal) graph $G=(V, E)$, where $V = \{x_1, \ldots, x_p\}$ with $\W$ being its adjacency matrix, i.e., $W_{ij}$ is the weight corresponding to the edge $(x_i, x_j) \in E$. For a given $x_j$ we define \textit{parent set} of $x_i$, denoted by $Pa(x_i)$, as the set of nodes $x_i$ for which $W_{ij} \neq  0$. This representation allows for a more straightforward causal semantics for the underlying interactions between the variables in the system.    

\subsection{Covariance Recovery}

Under the linear measurement system discussed previously, when the measurement noise is zero, the covariance of the observations $\y$ is given by 
\begin{equation}
\S_Y = \A\S\A^T.
\end{equation}
We further make the assumption that the covariance of $X$ is a sparse matrix. The covariance recovery can now be posed as the following convex program,
\begin{align*}
\min_{\S} & \quad \norm{\S}_1 \\
\text{subject to } & \quad \S_Y = \A\S\A^T. \tag{P\textsubscript{1}}
\end{align*}
Since we only have access to the observed samples of $\y$, the true covariance is approximated by the sample covariance, $\S_Y^{(N)} = (1/N)\sum_i \y_i\y_i^T$, and hence (P\textsubscript{1}) is relaxed as follows
\begin{align*}
\min_{\S} & \quad \norm{\S}_1 \\
\text{subject to } & \quad \norm{\S_{Y}^{(N)} - \A\S\A^T}_F^2 \leq \kappa. \tag{P\textsubscript{2}}
\end{align*}
Upon vectorization, we have $\A\S\A^T = (\A \otimes \A) \text{vec}(\S)$, where $\otimes$ denotes the Kronecker product. This gives the following equivalent formulation of (P\textsubscript2),
\begin{align*}
\min_{\S} & \quad \norm{\text{vec}(\S)}_1 \\
\text{subject to} & \quad \norm{\text{vec}(\S_{Y}^{(N)}) - (\A\otimes\A)\text{vec}(\S)}_2^2 \leq \kappa. 
\label{cp:cov_req}\tag{P\textsubscript{2}}
\end{align*}
In this vectorized form, $(\A\otimes \A)$ can be thought of as the new sensing matrix having a Kronecker product structure and $\text{vec}(\S)$ to be the sparse vector that has to be recovered.

\section{Design of Sensing Matrix using Density Evolution for Regular Recovery}
\label{sec:reg-sensing}

In this section, we describe the design scheme for the sensing matrix via the \textit{density evolution} methodology. For ease of notation let us denote $\bm{\gamma} = \text{vec}(\S_Y)$, $\bm{\chi} = \text{vec}(\S)$, and $\A^{\otimes} = \A\otimes\A$. The solution to (\ref{cp:cov_req}) can be viewed as the solution to the following \textit{maximum a posteriori} (MAP) estimator
\begin{equation}
\hat{\bm{\chi}} = \arg\max_{\bm{\c}} \exp\bigg(-\frac{\norm{\bm{\gamma} - \ak\bm{\c}}_2^2}{2\sigma^2}\bigg)\exp\big(-f(\bm{\c})\big), \label{eq:map}
\end{equation}
where $f(\bm{\c})$ is the generalized regularizer term. When $f(\bm{\c})$ is set to $\norm{\bm{\c}}_1$ then the MAP estimator is exactly equivalent to (\ref{cp:cov_req}). Here, we make a few assumptions on the sensing matrix and the regularizer: (i) The sensing matrix $\A$ is sparse with $EA_{ij} = 0$ and $A_{ij} \in \{0, \pm A^{-1/2}\}$, and (ii) The regularizer $f(\bm{\c})$ can be decomposed, $f(\bm{\c}) = \sum_{i} f(\c_i)$.

\begin{figure}[t]
     \centering
     \subfloat[][Block Factor Graph]{\includegraphics[width=0.20\linewidth]{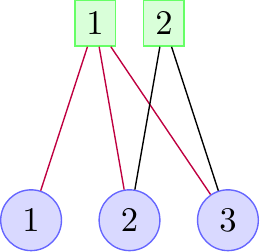}\label{fig:block_fg}}
     \hspace{1.5cm}
     \subfloat[][Complete Factor Graph]{\includegraphics[width=0.37\linewidth]{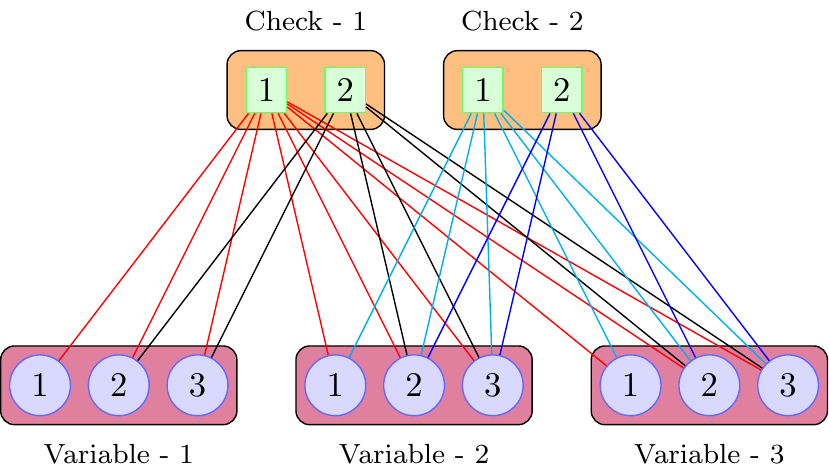} \label{fig:node_fg}}
     \caption{Illustration of the connections in factor graph corresponding to the Kronecker product when the sensing matrix is given by equation (\ref{eq:def_a}). (a) shows the connections at the block level, the number within the node corresponds to the block ID and as seen in (a), the connections at the block level are governed by $A$. (b) shows the connections at the node level.}
     \label{steady_state}
\end{figure}

To develop the density evolution framework, we associate (\ref{eq:map}) with a factor graph $\mathcal{G} = (\mathcal{V}, \mathcal{E})$ consisting of nodes corresponding to components of $\bm{\c}$ (variable nodes) and components of $\bm{\g}$ (check nodes), see Figure \ref{fig:node_fg}. An edge exists between $\c_i$ and $\g_j$ if $A^{\otimes}_{ij} \neq 0$. 

At this point it is important to illustrate some of the key structural properties of the factor graph that arises due to the Kronecker product nature of $\ak$, see Figures \ref{fig:block_fg} and \ref{fig:node_fg}. The check nodes and the variable nodes consist of $d$ and $p$ blocks respectively, and each check node block contains $d$ nodes and each variable node block contains $p$ nodes. $i$-th check node block is considered to be connected to $j$-th variable node block if any node in the $i$-th check node block is connected to any node in the $j$-th variable node block. This is true when $A_{ij} \neq 0$. The connection between the nodes in the $i$-th check node block and $j$-th variable node block, if it exists, is determined by $\A$. That is, within in the blocks, the $k$-th check node is connected to $l$-th variable node if $A_{kl} \neq 0$. Figure \ref{fig:node_fg} shows the factor graph for the following sensing matrix, 
\begin{equation}
\A = 
\begin{bmatrix}
1 & 1 & 1\\
0 & 1 & 1
\end{bmatrix}
\label{eq:def_a}
\end{equation}

\begin{figure}[b]
\centering
\includegraphics[width=0.3\linewidth]{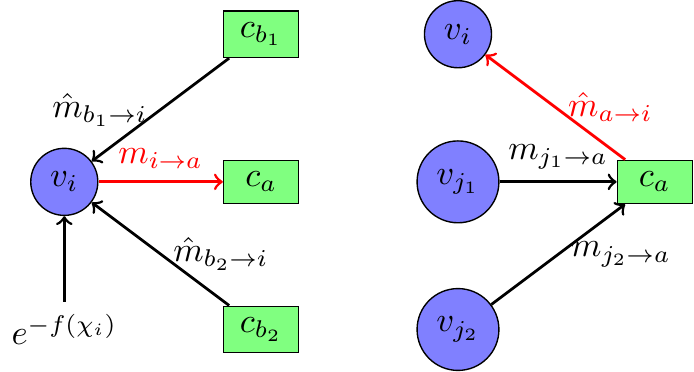}
\caption{Illustration of the flow of messages in the factor graph. The blue circles denote the variable nodes and the check nodes are depicted as the green rectangular nodes.}
\label{fig:messages}
\end{figure}

In view of the graphical model, recovery of $\S_X$ can be thought of as an inference problem over the factor graph which can be solved using the message-passing algorithm. Following the notations of \citet{hang_de}, let $m_{i\to a}^{(t)}$ denote the message going from the $i$-th variable node to the $a$-th check node at the $t$-th iteration. Similarly, let $\hat{m}_{a\to i}^{(t)}$ denote the message going from $a$-th check node to the $i$-th variable node at the $t$-th iteration, Figure \ref{fig:messages}. The message-passing algorithm is then given by

\begin{align}
m_{i\to a}^{(t+1)}(\c_i) \cong & e^{-f(\c_i)} \prod_{b\in\partial i\setminus a} \hat{m}_{b\to i}^{(t)}(\c_i);\\
\hat{m}_{a\to i}^{(t+1)}(\c_i) \cong & \int \prod_{j\in\partial a\setminus i} m_{j\to a}^{(t+1)}(\c_j) e^{-\frac{(\g_a - \Sigma_j A_{aj}\c_j)^2}{2\sigma^2}}d\c_j,
\end{align}
where $\partial a$, $\partial i$ denote the neighborhood of the $a$-th check node and the $i$-th variable node respectively and $\cong$ denotes equality up to a normalization constant. At iteration $t$, $\c_i$ can be recovered by taking argmax of the product of all the messages coming to the $i$-th variable node. 

To aid in the design of the sensing matrix, we define $\lambda(\alpha)$ and $\rho(\alpha)$ to be the distribution of the number of non-zero entries in the columns and rows of $\A$. The degree distribution of the check nodes and the variable nodes can then be obtained from $\lambda(\alpha)$ and $\rho(\alpha)$, refer to the appendix for more details. 

\subsection{Density Evolution}
\label{ssec:de-equal}

In order to design the sensing matrix using \textit{density evolution} (DE), the reconstruction of $\S$ has to be analyzed. To that end, the messages are treated as random variables, and in particular, they are chosen to be Gaussian distributed due to their simplicity. That is, $m_{i\to a}^{(t)} \sim \mathcal{N}(\mu_{i\to a}^{(t)}, v_{i\to a}^{(t)})$ and $\hat{m}_{a\to i}^{(t)} \sim \mathcal{N}(\hat{\mu}_{a\to i}^{(t)}, \hat{v}_{a\to i}^{(t)})$. To analyze the convergence of (\ref{eq:map}) we track the following two quantities
\begin{align}
E^{(t)} &= \frac{1}{d^2p^2}\sum_{a=1}^{d^2}\sum_{i=1}^{p^2}\Big(\mu_{i\to a}^{(t)} - \c_i\Big)^2;\\
V^{(t)} &= \frac{1}{d^2p^2}\sum_{a=1}^{d^2}\sum_{i=1}^{p^2}v_{i\to a}^{(t)}. 
\end{align}
Where $E^{(t)}$ and $V^{(t)}$ represent the average error and variance at iteration $t$. To enforce sparsity, the regularization function $f(\bm{\c})$ is set as $\beta\norm{\bm{\c}}_1$, this is equivalent to enforcing Laplacian prior on $\bm{\c}$. From DE analysis for the chosen prior, the average error and variance reduce to the following form, 
\begin{align}
E^{(t+1)} &= \textbf{E}_{\text{prior}(s)}\textbf{E}_{z\sim\mathcal{N}(0,1)}\bigg[\text{prox}\Big(s + a_1z\sqrt{E^{(t)}};
 \beta a_2V^{(t)}\Big) - s\bigg]^2\\ 
V^{(t+1)} &= \textbf{E}_{\text{prior}(s)}\textbf{E}_{z\sim\mathcal{N}(0,1)}\bigg[\beta a_2V^{(t)}\text{prox}^{\prime}\Big(s + a_1z\sqrt{E^{(t)}};
 \beta a_2V^{(t)}\Big)\bigg],
\end{align}
where $a_1$ is given by $\sum_{i, i^{\prime}, j, j^{\prime}} \rho_i\rho_{i^{\prime}}\lambda_j\lambda_{j^{\prime}}\sqrt{i i^{\prime}/ j j^{\prime}}$ and $a_2$ is given by $\sum_{i, i^{\prime}, j, j^{\prime}} \rho_i\rho_{i^{\prime}}\lambda_j\lambda_{j^{\prime}}(i i^{\prime}/ j j^{\prime})$. Also, prox$(a;b)$ denotes the soft-threshold function, and prox$^{\prime}(a;b)$ is the derivative of the soft-threshold function with respect to the first argument. For a detailed derivation of these quantities please refer to the appendix section. 

In designing the sensing matrix we would like to minimize the number of measurements needed to recover $\S$. We also need the message-passing algorithm to converge, i.e., $V^{(t)} \to 0$ and the average error should shrink to zero, $E^{(t)} \to 0$ as $t\to\infty$. However, enforcing $\lim_{t\to\infty} (E^{(t)}, V^{(t)}) = (0,0)$ is not straightforward and it requires running the DE updates numerically until convergence is achieved. For the case of sparse vector recovery, \cite{hang_de} showed that these requirements can be reduced to two inequality constraints making it easier to check for satisfiability. We extend this to the case of covariance recovery in the form of the following theorem.

\begin{theorem}
Let $\S$ be $k^2$-sparse and set $\beta$ to be $p^2/(c_0 \log(p/k))$ for $c_0 > 0$. Then, the necessary condition for $\lim_{t\to\infty} (E^{(t)}, V^{(t)}) = (0,0)$ results in $a_1^2 \leq p^2/k^2$ and $a_2 \leq p^2/(2c_0k^2\log(p/k))$, where $a_1 = \sum_{i, i^{\prime}, j, j^{\prime}} \rho_i\rho_{i^{\prime}}\lambda_j\lambda_{j^{\prime}}\sqrt{i i^{\prime}/ j j^{\prime}}$ and $a_2 = \sum_{i, i^{\prime}, j, j^{\prime}} \rho_i\rho_{i^{\prime}}\lambda_j\lambda_{j^{\prime}}(i i^{\prime}/ j j^{\prime})$.  
\end{theorem}
Therefore the design of the sensing matrix can be posed as the following optimization problem, 
\begin{align}
\min_{\substack{\bm{\lambda}\in\Delta_{d_v};\\ \bm{\rho}\in\Delta_{d_c}}} & \quad\frac{d}{p} = \frac{\sum_{i\geq 2} i\lambda_i}{\sum_{j \geq 2} j\rho_j} \label{eq:sense}\\
\text{s.t} & \quad a_1^2 \leq \frac{p^2}{k^2}\\
	& a_2 \leq \frac{p^2}{2c_0k^2\log(p/k)}\\
	& \lambda_1 = \rho_1 = 0, \label{eq:owmp}
\end{align}
where $\Delta_d$ is a d-dimensional simplex, $d_v$ and $d_c$ denote the maximum column and row degree respectively of sensing matrix $\A$. The final constraint (\ref{eq:owmp}) is added to avoid one-way message passing. Once we have the distributions $\bm{\lambda}$ and $\bm{\rho}$ we then sample the sensing matrix such that the number of non-zero entries in the rows and columns satisfies the obtained distributions. For every non-zero entry of $\A$, $P(A_{ij} = A^{-1/2}) = P(A_{ij} = -A^{-1/2}) = \frac{1}{2}$. With the sensing matrix obtained, (\ref{cp:cov_req}) can be solved using any convex program solver.

\section{Sensing Matrix for Preferential Covariance Recovery}
\label{sec:pref-req}
In this section, we extend the density evolution based sensing matrix design to the case of preferential recovery of the covariance matrix. That is, we employ the DE framework to construct sensing matrices that provide higher importance to a sub-block of the covariance matrix. In other words, we treat certain variables as important and try to recover the covariance between the important variables with higher accuracy. 

\subsection{Density Evolution}
\label{ssec:de-unequal}

The unknown signal $\x$ is divided into two parts $\x_H \in \R^{n_H}$ (high priority), and $\x_L \in \R^{n_L}$ (low priority) and without loss of generality we assume that $\x = (\x_H, \x_L)$. This splits the covariance into four sub-matrices,
\begin{equation}
    \S_X =  
    \begin{bmatrix}
        \S_{HH} & \S_{HL}\\
        \S_{LH} & \S_{LL}
    \end{bmatrix}.
\end{equation}
In this case, we would like to place higher importance on $\S_{HH}$ and design the sensing matrix in order to recover the higher priority sub-block with higher accuracy than the other components. To that end, we introduce the degree distributions $\lambda_H(\alpha)$ and $\lambda_L(\alpha)$ corresponding to the first $n_H$ columns and the last $n_L$ columns of the sensing matrix respectively. Similarly, $\rho_H(\alpha)$ and $\rho_L(\alpha)$ correspond to the degree distribution of the first $n_H$ rows and the last $n_L$ rows of the sensing matrix.

Generalizing the analysis for regular sensing, the average error and the variance for each sub-matrix of $\S_X$ are separately tracked. For $\S_{HH}$ sub-block, $E_{HH}$ is defined as $\sum_{a}\sum_{i\in HH}(\mu_{i\to a} - \chi_i)^2/(d^2\cdot n_H^2)$ and $V_{HH} = \sum_{a}\sum_{i\in HH}v_{t\to a}/(d^2\cdot n_H^2)$. The average error and variance for LH, HL, and LL is defined in a similar manner. Similar to regular sensing by assuming a Laplacian prior on $\bm{\c}$ we then have
\begin{align}
E_{HH}^{(t+1)} &= \textbf{E}_{\text{prior}(s)}\textbf{E}_{z\sim\mathcal{N}(0,1)}\bigg[\text{prox}\Big(s + zb_{HH, 1}^{(t)}; b_{HH, 2}^{(t)}\Big) - s\bigg]^2;\\ 
V_{HH}^{(t+1)} &= \textbf{E}_{\text{prior}(s)}\textbf{E}_{z\sim\mathcal{N}(0,1)}\bigg[b_{HH, 2}^{(t)}\text{prox}^{\prime}\Big(s + zb_{HH, 1}^{(t)};  b_{HH, 2}^{(t)}\Big)\bigg],
\end{align}
where $b_{HH, 1}^{(t)}$ and $b_{HH, 2}^{(t)}$ are defined as follows
\begin{align}
   b_{HH, 1}^{(t)} &= \sum_{\ell\ell^{\prime}, ii^{\prime}, jj^{\prime}, kk^{\prime}} \lambda_{H, \ell}\lambda_{H, \ell^{\prime}}\rho_{H, i}\rho_{H, i^{\prime}}\rho_{H, j}\rho_{L, j^{\prime}}\rho_{L, k}\rho_{L, k^{\prime}} \sqrt{\frac{A\sigma^2 + ii^{\prime}E_{HH}^{(t)} + jj^{\prime}E_{HL}^{(t)} + kk^{\prime}E_{LL}^{(t)}}{\ell\ell^{\prime}}}; \\
      b_{HH, 2}^{(t)} &= \sum_{\ell\ell^{\prime}, ii^{\prime}, jj^{\prime}, kk^{\prime}} \lambda_{H, \ell}\lambda_{H, \ell^{\prime}}\rho_{H, i}\rho_{H, i^{\prime}}\rho_{H, j}\rho_{L, j^{\prime}}\rho_{L, k}\rho_{L, k^{\prime}}\frac{A\sigma^2 + ii^{\prime}V_{HH}^{(t)} + jj^{\prime}V_{HL}^{(t)} + kk^{\prime}V_{LL}^{(t)}}{\ell\ell^{\prime}}.
      \label{eq;pref-req-param}
\end{align}
For the case of preferential sensing, the sensing matrix must satisfy the following constraints.
\begin{enumerate}[label=Req \arabic*.]
    \item We require consistency with respect to the number of non-zero entries in the sensing matrix. Starting with the high priority part, the number of non-zero entries in the first $n_H$ columns is given by $n_H(\sum_i\lambda_{H,i})$ (counting the non-zeros by column) and $d(\sum_ii\rho_{H,i})$ (counting by rows). Therefore we have the following constraint
    $$n_H\bigg(\sum_i\lambda_{H,i})\bigg) = d\bigg(\sum_i\rho_{H,i}\bigg).$$
    Similarly, the consistency requirement on the low-priority part would yield $n_L(\sum_i\lambda_{L,i}) = d(\sum_ii\rho_{H,i})$. 

    \item We require the variances to converge to zero. That is,
    $$\lim_{t\to\infty} \Big(V_{HH}^{(t)}, V_{HL}^{(t)}, V_{LL}^{(t)}\Big)  = (0,0,0)$$
    This implies that the message-passing algorithm on the factor graph converges. Here we exclude $V_{LH}$ due to the symmetric nature of the covariance matrix. 
    \item Due to the preferential nature of the design we require that the error in the high-priority part of the covariance is lower than the other sub-matrices. In other words, let $\delta_{E, HH}^{(t)} = E_{HH}^{(t+1)} - E_{HH}^{(t)}$, and we similarly define $\delta_{E, HL}^{(t)}$ and $\delta_{E, LL}^{(t)}$, we want $|\delta_{E, HH}^{(t)}| \leq |\delta_{E, HL}^{(t)}|$ and $|\delta_{E, HH}^{(t)}| \leq |\delta_{E, LL}^{(t)}|$ for all $t \geq T_0$ for some $T_0$. 
\end{enumerate}
Hence, the design of the sensing matrix can be posed as the following convex problem, 
\begin{align}
\min_{\substack{\lambda_H\in\Delta_{d_{v_H}};\\ \lambda_L\in\Delta_{d_{v_L}};\\\rho_{H}\in\Delta_{d_{c_H}};\\\rho_{L}\in\Delta_{d_{c_L}}}} & \quad\frac{d}{p} = \frac{n_L\sum_{i} i\lambda_{L,i} + n_H\sum_ii\lambda_{H,i}}{\sum_{j} j(\rho_{H,j} + \rho_{L,j})} \label{eq:sense}\\
\text{s.t} & \quad \frac{\sum_ii\lambda_{L,i}}{\sum_ii\lambda_{H,i}} \times \frac{\sum_ii\rho_{H,i}}{\sum_ii\rho_{L,i}} = \frac{n_H}{n_L};\\
	& \text{Requirement 2 \& 3};\\
	& \lambda_{H,1} = \lambda_{L,1} = \rho_{H,1} = \rho_{L,1} = 0, \label{eq:owmp-pref}
\end{align}
\subsection{Constraint Relaxation for Laplacian Prior}
\label{ssec:const-relax}

Consider a sparse covariance matrix where the high priority subpart is $k_{HH}$-sparse and the low priority subpart is $k_{LL}$-sparse, with the added assumption that $k_{HH}/n_h \gg k_{LL}/n_L$. As stated in section \ref{ssec:de-equal}, directly enforcing requirements 2 and 3 in equation (24) is not straightforward. Fortunately, by assuming the prior to be Laplacian, requirements 2 and 3 can be relaxed to obtain the following inequalities constraints that are convex in the degree polynomials 
\begin{multline}
    \Bigg\{\bigg[\frac{\beta_{HH}k_{HH}}{n_{HH}}\bigg(\sum_{\ell}\frac{\lambda_{H, \ell}}{\ell}\bigg)^2\bigg] + 2\bigg[\frac{\beta_{HL}k_{HL}}{n_{HL}}\sum_{\ell, k}\frac{\lambda_{H, \ell}\lambda_{L,k}}{\ell k}\bigg] + \bigg[\frac{\beta_{LL}k_{LL}}{n_{LL}}\bigg(\sum_{\ell}\frac{\lambda_{L, \ell}}{\ell}\bigg)^2\bigg]\Bigg\}\\
    \times \Bigg[\bigg(\sum_{i}i\rho_{H,i}\bigg)^2 + \bigg(\sum_{i}i\rho_{L,i}\bigg)^2\Bigg]^2 \leq 1.
\label{eq:req2}
\end{multline}
\begin{align}
    \sqrt{\frac{k_{HH}}{n_{HH}}}\Bigg(\sum_{\ell}\frac{\lambda_{H,\ell}}{\sqrt{\ell}}\Bigg) &\leq \sqrt{\frac{k_{HL}}{n_{HL}}}\Bigg(\sum_{\ell}\frac{\lambda_{L,\ell}}{\sqrt{\ell}}\Bigg);\\
    \bigg(\frac{k_{HH}}{n_{HH}}\bigg)^{1/4}\Bigg(\sum_{\ell}\frac{\lambda_{H,\ell}}{\sqrt{\ell}}\Bigg) &\leq \bigg(\frac{k_{LL}}{n_{LL}}\bigg)^{1/4}\Bigg(\sum_{\ell}\frac{\lambda_{L,\ell}}{\sqrt{\ell}}\Bigg).
    \label{eq:req3-2}
\end{align}
Here, equation (\ref{eq:req2}) corresponds to requirement 2 and equation (\ref{eq:req3-2}) corresponds to requirement 3. The above inequalities are convex with respect to the degree polynomials and hence can be solved using any convex program solver. The key idea behind the relaxation is to approximate $\delta_{E, HH}^{(t)}, \delta_{V, HH}^{(t)}$, $\delta_{E, HL}^{(t)}, \delta_{V, HL}^{(t)}$ and $\delta_{E, LL}^{(t)}, \delta_{V, LL}^{(t)}$ by its first-order Taylor series approximation and enforcing the operator norm of the Jacobian to be less than one, readers are referred to the appendix for more details.     

\section{Graph Structure Recovery}
\label{sec:graph-struct-rec}

In this section, we discuss the steps involved in the recovery of the weighted adjacency matrix $\W$ encoding the underlying graph structure of $\x$. We first estimate the covariance matrix of $\x$, $\S_X$ from the compressed measurements obtained using the sensing matrix designed via the density evolution objective. Using the estimated covariance matrix, any consistent causal discovery method can be used to infer the underlying graph structure. In our case, we assume that the intrinsic noise variables are i.i.d Gaussian and that the graph $G$ is acyclic (Gaussian Bayesian Network, GBN). The graph structure is then recovered using the algorithm developed by \citet{ligbn}. Once the covariance matrix is retrieved as discussed in the previous section, the precision matrix is obtained using \textit{Constrained $\ell_1$-minimization for Inverse Matrix Estimation }(CLIME), a constrained convex optimization framework, proposed by \citet{clime}. CLIME forces the precision matrix to approximate the inverse of the estimated covariance matrix by minimizing $\norm{\hat{\S}\bm{\Theta} - \bm{I}}_{\infty}$. Using the estimated covariance and precision matrix, the following steps are performed to obtain the structure of the GBN:
\begin{enumerate}
\item Identify the Markov blanket of each node ($MB_i$). This is done by looking at indices of the non-zero entries of each column/row of the precision matrix $\bm{\Omega}$.
\item Compute the regression coefficients ($\bm{\theta_i}$), which depend on the covariance matrix and the Markov blankets. The regression coefficients are defined as,
$\bm{\theta_i^T}\x_{-i} = \textbf{E}[X_i | X_{-i} = \x_{-i}].$
\item Identify the terminal nodes, which depend on the precision matrix $\bm{\Omega}$ and regression coefficients $\bm{\theta}_i$. Let us define, 
$r_i = \max_{j \in MB_i} \bigg|\frac{\Omega_{ij}}{\theta_{ij}}\bigg|.$
Then $v = \arg\min_i r_i$ is the terminal node. Once we have the terminal node, the Markov blanket gives the parents of the terminal node. 
\item The terminal node is removed and the joint distribution is marginalized with respect to the terminal node.  
\end{enumerate}
These four steps are repeated until only one node is left in the graph providing us with all the parent-child relations in the graph and thereby the structure of the Bayesian network. 

\section{Experiments}
\label{sec:experiments}
\begin{figure}[t]
\centering
\includegraphics[width=\linewidth]{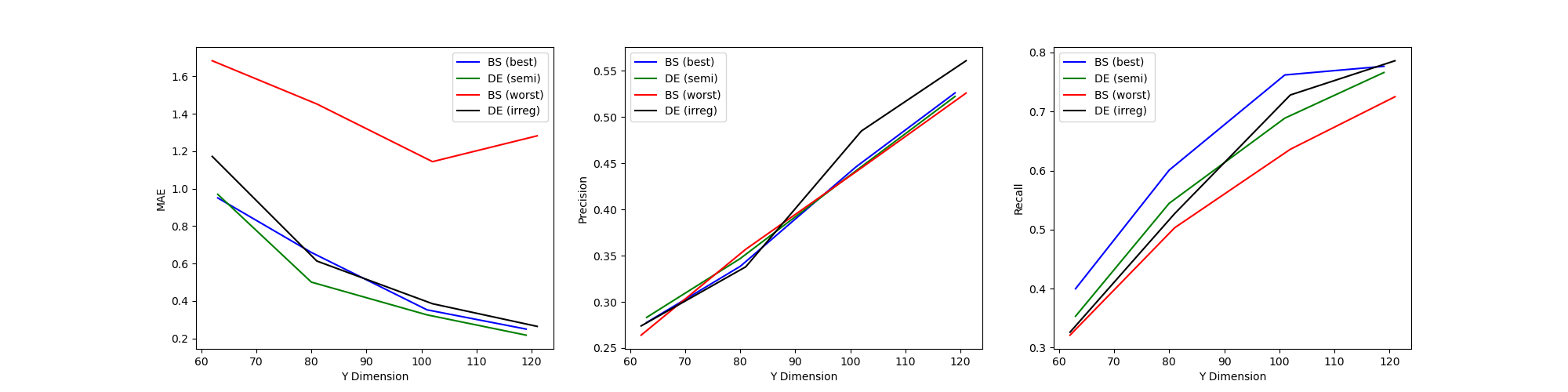}
\caption{Comparision of the performance of the proposed regular sensing systems with that of \citet{mksetch}, denoted as BS in the plots. For the baseline, we chose two different versions of the sensing system, one where their hyperparameters are tuned (BS - best) and one where the parameters were initialized randomly (BS - worst). The figure shows the performance when the number of nodes in the graph $p = 200$.}
\label{fig:cov_comp_200}
\end{figure}

\begin{figure}[t]
    \centering
    \includegraphics[width=\linewidth]{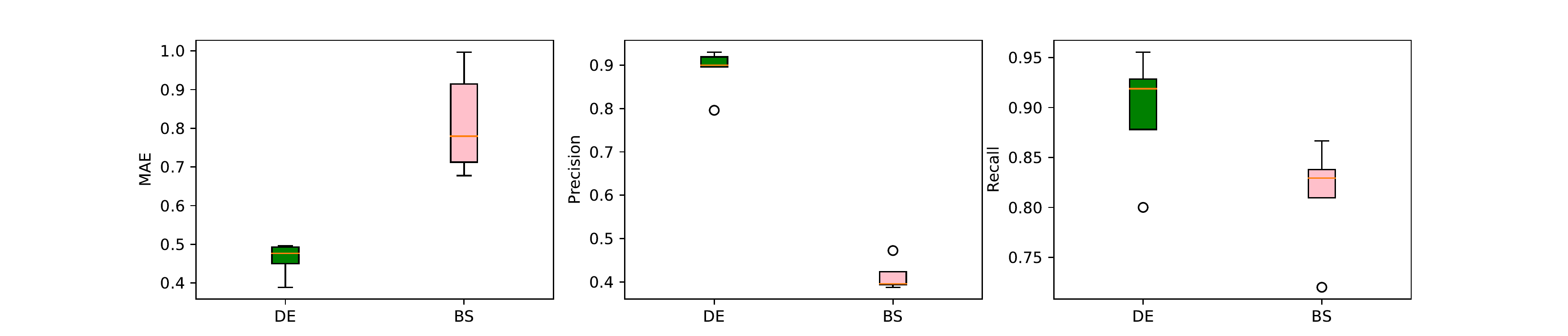}
    \caption{Performance comparison on the covariance recovery task between the proposed preferential sensing matrix (denoted as DE) and that of \citet{mksetch}. Here the number of nodes set to $d=200$, with the first 50 nodes corresponding to the high-priority part of the covariance. The accuracy is measured over the high-priority portion of the covariance matrix.}
    \label{fig:cov-pref-req}
\end{figure}
In this section, we present the numerical experiments performed to evaluate covariance and graph recovery. To generate the GBN, we sampled directed graphs from Erd\"os-R\'enyi class of random graphs with edge weights set to $\pm1/2$ with probability $1/2$. We first study the effectiveness of the sensing system for recovery of the entire covariance recovery matrix followed by preferential recovery of the high-priority portion of the covariance matrix. We compare the performance with the current state-of-the-art \citep{mksetch}, where the sensing matrix is the adjacency matrix of $\delta$-left-regular bipartite graph. We then evaluate the performance of the sensing system for graph structure recovery. 

\subsection{Covariance Recovery}

\subsubsection{Regular Sensing}

we consider three different design schemes for constructing the sensing matrix. (i) \textit{Fixed row degree and variable column degree}. In this case, $\rho_i = 1$ when $i = d_c$ and 0 otherwise. We then solve (\ref{eq:sense}) for $\bm{\lambda}$, (ii) \textit{Fixed column degree and variable row degree}. In this case, $\lambda_i = 1$ when $i = d_v$ and 0 otherwise. Equation (\ref{eq:sense}) is then solved for $\bm{\rho}$, and (iii) \textit{Variable row and column degree}. In this case we solve (\ref{eq:sense}) for both $\bm{\lambda}$ and $\bm{\rho}$. 
In cases (1) and (2), the resulting optimization program is readily solvable by any convex program solver. For case (3), we first keep $\bm{\lambda}$ constant and solve for $\bm{\rho}$, then using the obtained solution for $\bm{\rho}$ we solve for $\bm{\lambda}$. 

The recovery performance is evaluated using three metrics, namely, (1) \textit{Maximum Absolute Error} (MAE) which is given by the maximum absolute difference between the estimate covariance matrix and the ground truth covariance matrix (lower the better), (2) \textit{Precision} of the recovery of the support of the covariance matrix, since the covariance is sparse we measure the percentage of estimated support that belongs to the support of the ground truth covariance (higher the better), and (3) \textit{Recall} which measures the percentage of the support of the ground truth covariance that has been recovered (higher the better).

The three design schemes attain similar performance with respect to all the metrics, as seen in Figure \ref{fig:cov_comp_200}. Hence there isn't any inherent advantage of choosing one over the other. We can also observe that the density evolution based sensing matrices achieve similar performance to that of \citet{mksetch} when $\delta$ is tuned. On the other hand, improper assignment of $\delta$ results in poor performance compared to the density evolution based design. 

\subsubsection{Preferential Sensing}

For the case of preferential sensing, we considered graphs with $p=200$ nodes, where we choose covariance between the first $n_H=50$ nodes to be of higher priority. The measurements are then compressed down to $d=60$ dimensions. The performance of the preferential sensing matrix is compared with that of \citet{mksetch} with respect to the same metrics described in the previous section on the high-priority sub-matrix of the covariance. In this case, we fix the degree distribution of the check nodes and solve for the degree distribution of the variable using the procedure described in section \ref{ssec:de-unequal}. As seen from Figure \ref{fig:cov-pref-req}, the proposed preferential sensing matrix outperforms the baseline with respect to all the error metrics, showcasing that the density evolution framework can be used to design sensing matrices that are capable of providing preferential treatment to a portion of the full covariance matrix. 
\begin{figure}[t]
\centering
\includegraphics[width=0.8\linewidth]{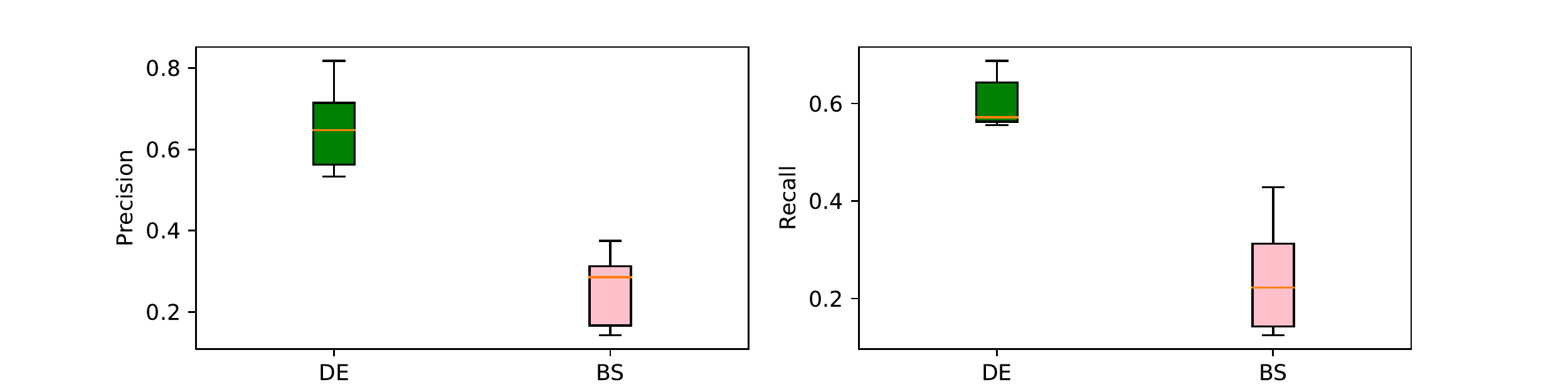}
\caption{Comparison of the performance of the proposed preferential sensing system with that of \cite{mksetch} (BS - best, BS - worst lines in the plots). Number of nodes, $p = 200$. The performance is evaluated with respect to \textit{precision} and \textit{recall} of the edges in the graph subset to the high-priority nodes.}
\label{fig:graph-comp-pref-req}
\end{figure}

\subsection{Graph Structure Recovery}
\label{ssec:res-graph-rec}

Using the covariance matrix recovered from the observations $\y$, CLIME \citep{clime} was used to estimate the precision matrix. The graph structure is then recovered using the covariance and the precision matrix as described in section \ref{sec:graph-struct-rec}. The performance is evaluated using precision and recall as metrics. For preferential recovery, we only consider the edges connecting the high-priority nodes for evaluating the performance. The proposed regular sensing matrix achieves similar performance to that of the baseline, like in the case of covariance recovery and hence we refer the readers to the appendix for details.  Figure \ref{fig:graph-comp-pref-req} shows the performance comparison between the proposed preferential sensing matrix and the baseline. As seen from the figure, we see a similar trend to that of covariance recovery, i.e., the preferential sensing system outperforms the baseline with respect to all the metrics. This shows that having a preferential sensing scheme does indeed help with recovering a part of the graph structure (that is of interest) more accurately.  

\section{Conclusion}
\label{sec:conclusion}

In this paper, we presented a general framework for collecting lower dimensional samples of the signal generated from a GBN for accurate recovery of the covariance and graph structure under (i) regular and (ii) preferential sensing regimes. We also showcased the feasibility of our approach through numerical simulations. There are several directions that could be of interest in the future. While we restricted our focus to GBNs, exploring other types of additive noise distributions would be an interesting avenue. The proposed density evolution framework can also be extended to support other types of prior on the covariance matrix, like low-rank. 


\bibliography{example_paper}

\begin{thebibliography}{}

\bibitem[Berinde et~al., 2008]{binary_vec}
Berinde, R., Gilbert, A.~C., Indyk, P., Karloff, H., and Strauss, M.~J. (2008).
\newblock Combining geometry and combinatorics: A unified approach to sparse
  signal recovery.
\newblock In {\em 2008 46th Annual Allerton Conference on Communication,
  Control, and Computing}, pages 798--805. IEEE.

\bibitem[Bollen, 1989]{sem1}
Bollen, K.~A. (1989).
\newblock {\em Structural equations with latent variables}, volume 210.
\newblock John Wiley \& Sons.

\bibitem[Cai et~al., 2011]{clime}
Cai, T., Liu, W., and Luo, X. (2011).
\newblock A constrained $\ell_1$ minimization approach to sparse precision
  matrix estimation.
\newblock {\em Journal of the American Statistical Association},
  106(494):594--607.

\bibitem[Candes et~al., 2006]{cs1}
Candes, E., Romberg, J., and Tao, T. (2006).
\newblock Robust uncertainty principles: exact signal reconstruction from
  highly incomplete frequency information.
\newblock {\em IEEE Transactions on Information Theory}, 52(2):489--509.

\bibitem[Chickering et~al., 2004]{np-hard}
Chickering, M., Heckerman, D., and Meek, C. (2004).
\newblock Large-sample learning of bayesian networks is np-hard.
\newblock {\em Journal of Machine Learning Research}, 5.

\bibitem[Dasarathy et~al., 2015]{mksetch}
Dasarathy, G., Shah, P., Bhaskar, B.~N., and Nowak, R.~D. (2015).
\newblock Sketching sparse matrices, covariances, and graphs via tensor
  products.
\newblock {\em IEEE Transactions on Information Theory}, 61(3):1373--1388.

\bibitem[DeVore, 2007]{cs2}
DeVore, R.~A. (2007).
\newblock Deterministic constructions of compressed sensing matrices.
\newblock {\em Journal of complexity}, 23(4-6):918--925.

\bibitem[Donoho et~al., 2009]{mpcs}
Donoho, D.~L., Maleki, A., and Montanari, A. (2009).
\newblock Message-passing algorithms for compressed sensing.
\newblock {\em Proceedings of the National Academy of Sciences},
  106(45):18914--18919.

\bibitem[Gallager, 1962]{ldpc}
Gallager, R. (1962).
\newblock Low-density parity-check codes.
\newblock {\em IRE Transactions on Information Theory}, 8(1):21--28.

\bibitem[Ghoshal and Honorio, 2017]{ligbn}
Ghoshal, A. and Honorio, J. (2017).
\newblock Learning identifiable gaussian bayesian networks in polynomial time
  and sample complexity.
\newblock In Guyon, I., Luxburg, U.~V., Bengio, S., Wallach, H., Fergus, R.,
  Vishwanathan, S., and Garnett, R., editors, {\em Advances in Neural
  Information Processing Systems}, volume~30. Curran Associates, Inc.

\bibitem[Guo et~al., 2020]{survey}
Guo, R., Cheng, L., Li, J., Hahn, P.~R., and Liu, H. (2020).
\newblock A survey of learning causality with data: Problems and methods.
\newblock {\em ACM Computing Surveys (CSUR)}, 53(4):1--37.

\bibitem[Kaplan et~al., 2018]{nogauss}
Kaplan, A., Pohl, V., and Lee, D.~G. (2018).
\newblock On compressive sensing of sparse covariance matrices using
  deterministic sensing matrices.
\newblock In {\em 2018 IEEE International Conference on Acoustics, Speech and
  Signal Processing (ICASSP)}, pages 4019--4023. IEEE.

\bibitem[Krzakala et~al., 2012a]{statpcs2}
Krzakala, F., M{\'e}zard, M., Sausset, F., Sun, Y., and Zdeborov{\'a}, L.
  (2012a).
\newblock Probabilistic reconstruction in compressed sensing: algorithms, phase
  diagrams, and threshold achieving matrices.
\newblock {\em Journal of Statistical Mechanics: Theory and Experiment},
  2012(08):P08009.

\bibitem[Krzakala et~al., 2012b]{statpcs1}
Krzakala, F., M{\'e}zard, M., Sausset, F., Sun, Y., and Zdeborov{\'a}, L.
  (2012b).
\newblock Statistical-physics-based reconstruction in compressed sensing.
\newblock {\em Physical Review X}, 2(2):021005.

\bibitem[Mezard and Montanari, 2009]{bethe}
Mezard, M. and Montanari, A. (2009).
\newblock {\em Information, physics, and computation}.
\newblock Oxford University Press.

\bibitem[M{\"u}ller et~al., 2008]{comp_moti}
M{\"u}ller, J., Kuttler, C., and Hense, B.~A. (2008).
\newblock Sensitivity of the quorum sensing system is achieved by low pass
  filtering.
\newblock {\em Biosystems}, 92(1):76--81.

\bibitem[Pearl, 1988]{pearl}
Pearl, J. (1988).
\newblock {\em Probabilistic reasoning in intelligent systems: networks of
  plausible inference}.
\newblock Morgan kaufmann.

\bibitem[Pearl, 2009]{sem2}
Pearl, J. (2009).
\newblock {\em Causality}.
\newblock Cambridge University Press, 2 edition.

\bibitem[Spirtes et~al., 2000]{pcalgo}
Spirtes, P., Glymour, C.~N., Scheines, R., and Heckerman, D. (2000).
\newblock {\em Causation, prediction, and search}.
\newblock MIT press.

\bibitem[Van~de Geer and B{\"u}hlmann, 2013]{score}
Van~de Geer, S. and B{\"u}hlmann, P. (2013).
\newblock $\ell_0$-penalized maximum likelihood for sparse directed acyclic
  graphs.
\newblock {\em The Annals of Statistics}, 41(2):536--567.

\bibitem[Zdeborov{\'a} and Krzakala, 2016]{statpcs3}
Zdeborov{\'a}, L. and Krzakala, F. (2016).
\newblock Statistical physics of inference: Thresholds and algorithms.
\newblock {\em Advances in Physics}, 65(5):453--552.

\bibitem[Zhang et~al., 2022]{hang_de}
Zhang, H., Abdi, A., and Fekri, F. (2022).
\newblock a general compressive sensing construct using density evolution.
\newblock {\em IEEE Transactions on Signal Processing}, pages 1--16.

\bibitem[Zheng et~al., 2018]{notears}
Zheng, X., Aragam, B., Ravikumar, P.~K., and Xing, E.~P. (2018).
\newblock Dags with no tears: Continuous optimization for structure learning.
\newblock In Bengio, S., Wallach, H., Larochelle, H., Grauman, K.,
  Cesa-Bianchi, N., and Garnett, R., editors, {\em Advances in Neural
  Information Processing Systems}, volume~31. Curran Associates, Inc.

\end{thebibliography}
\newpage
\appendix

\section{Degree Distribution of Check nodes and Variable nodes}

\begin{figure}[h]
\centering
\includegraphics[width=0.4\linewidth]{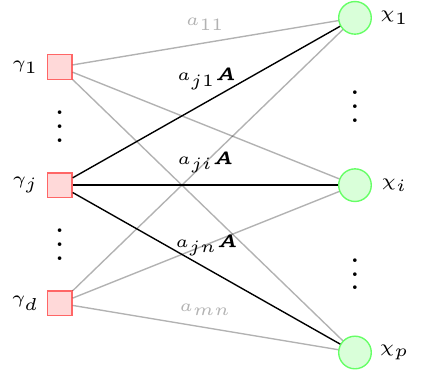}
\caption{Illustration of the connections in the factor graph at block level.}
\label{fig:block_connections}
\end{figure}

\begin{figure}[h]
\centering
\includegraphics[width=0.4\linewidth]{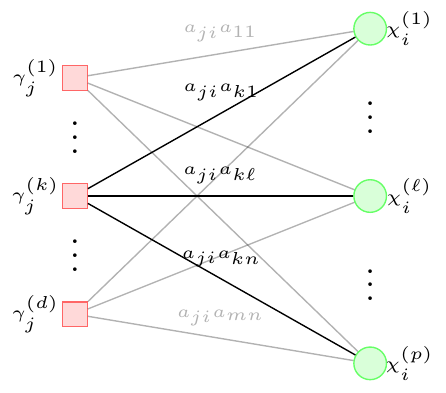}
\caption{Illustration of the connections between blocks $\g_j$ and $\c_i$}
\label{fig:node_connections}
\end{figure}

As described in section 3.1, let $\bm{\lambda} \in \Delta_{d_v}$, $\bm{\rho} \in \Delta_{d_c}$ be the degree distributions of columns and rows of $\A$. We can divide $\bm{\g}$ and $\bm{\c}$ into blocks of size $d$ and $p$ nodes respectively. Each block corresponds to a column of $\S_Y$ and $\S$. Let $\g_i$ denote the $i$-th block of $\bm{\g} $ and similarly let $\c_j$ denote the $j$-th block of $\bm{\c}$. In the factor graph, \ref{fig:node_fg}, blocks $\g_i$ and $\c_j$ are connected if at least one node in $\g_i$ is connected to at least one node in $\c_j$. The connections at the block level are defined by the sensing matrix $\A$. In other words, $\g_i$ and $\c_j$ are connected if $A_{ij} \neq 0$. Figure \ref{fig:block_connections}, illustrates the connections at the block level. 
Let us now focus on the connections between the nodes in block $\g_j$ and $\c_i$. We denote $\g_j^{(k)}$ to be the $k$-th node in check node block $j$ and $\c_i^{\ell}$ to be the $\ell$-th node in the variable node block $i$. The connections between the blocks $\g_j$ and $\c_i$, if it exists ($A_{ji} \neq 0$), is again characterized by $\A$. Figure \ref{fig:node_connections} illustrated the connected between the nodes in a variable node block and a check node block. 
Therefore we now have,
\begin{equation}
\g_j^{(k)} = \sum_{i=1}^p\sum_{\ell=1}^p A_{ji}A_{k\ell} \c_i^{(\ell)}.
\end{equation}
Since deg$(\g_j^{(k)})$ would be the number of non-zero terms in the above summation, we then have deg$(\g_j^{(k)}) = \text{deg}(A^{(j)})\text{deg}(A^{(k)})$, where $A^{(j)}$ denotes the $j$-th row of $\A$. Using a similar argument we can also conclude that deg$(\c_i^{(\ell)}) = \text{deg}(A_{i})\text{deg}(A_{\ell})$, where $A_i$ denotes the $i$-th column of $\A$. Since deg$(A_i) \in \{1,\ldots, d_v\}$ and deg$(A^{(j)}) \in \{1, \ldots, d_c\}$ we have that deg$(\g_j^{(k)}) \in \{1, \ldots, d_c^2\}$ and deg$(\c_i^{(\ell)}) = \{1, \ldots, d_v^2\}$. Therefore we have 
\begin{equation}
P\Big(\text{deg}(\g_j^{(k)}) = k\Big) = \sum_{j, j^{\prime} : jj^{\prime} = k} \rho_j\rho_{j^{\prime}}
\end{equation}
And, 
\begin{equation}
P\Big(\text{deg}(\c_i^{(\ell)}) = k\Big) = \sum_{i, i^{\prime} : ii^{\prime} = k} \lambda_i\lambda_{i^{\prime}}
\end{equation}

\section{Derivation of DE Update Equations}
\label{sec:DE-update}
As described in section 3, in order to analyze the convergence of the message-passing algorithm, the two quantities given by equations (8) and (9) are tracked over the course of the algorithm, re-written here for convenience. 
\begin{align*}
E^{(t)} &= \frac{1}{d^2p^2}\sum_{a=1}^{d^2}\sum_{i=1}^{p^2}\Big(\mu_{i\to a}^{(t)} - \c_i^{*}\Big)^2;\\
V^{(t)} &= \frac{1}{d^2p^2}\sum_{a=1}^{d^2}\sum_{i=1}^{p^2}v_{i\to a}^{(t)}. 
\end{align*}
To simplify these two quantities, we need to simplify the messages flowing through the factor graph. To that end, we start with the messages sent from the check nodes to the variable nodes, $\hat{m}_{a\to i}^{(t)} \sim \mathcal{N}\Big(\hat{\mu}_{a\to i}^{(t)}, \hat{v}_{a\to i}^{(t)}\Big)$. \cite{hang_de} derived a simplified update for the $\hat{\mu}_{a\to i}^{(t)}$ and $\hat{v}_{a\to i}^{(t)}$ in Lemma 6. Here we list the lemma and modify it our purpose to account for the Kronecker product sensing matrix. 

\begin{lemma}
Consider the message flowing from check node $a$ to variable node $i$, $\hat{m}_{a\to i}^{(t)} \sim \mathcal{N}\Big(\hat{\mu}_{a\to i}^{(t)}, \hat{v}_{a\to i}^{(t)}\Big)$. Then the following update can be obtained at the $(t+1)$-th iteration. 
\begin{align}
\hat{\mu}_{a\to i}^{(t+1)} &= \c_i + A \sum_{j\in \partial a\setminus i} A_{ai}^{\otimes}A_{aj}^{\otimes}\Big(\c_j - \mu_{j\to a}^{(t)}\Big) + AA_{ai}^{\otimes}n_a;\\
\hat{v}_{a\to i}^{(t+1)} &= A\sigma^2 + |\partial a|V^{(t)}.
\end{align}
Where $\c_i$ is the $i$-th variable node and $|\partial a|$ is the degree of the check node $a$.  
\end{lemma}

Now consider the message going from variable nodes to check nodes, $m_{i\to a}^{(t)} \sim \mathcal{N}\Big(\mu_{i\to a}^{(t)}, v_{i\to a}^{(t)}\Big)$. Using the previous lemma and exploiting some properties of Gaussian distribution with some approximations along the way, $\mu_{i\to a}^{(t)}$ and $v_{i\to a}^{(t)}$ can be updated as follows, here we also make use of the characterization of degrees of check nodes and the variable nodes from the section A. The readers are referred to \cite{hang_de} for more details. 
\begin{align}
\mu_{i\to a}^{(t+1)} &\approx h_{\text{mean}}\Bigg(\c_i + z\sum_{i,i^{\prime}, j, j^{\prime}} \rho_i\rho_{i^{\prime}}\lambda_j\lambda_{j^\prime} \sqrt{\frac{ii^{\prime}E^{(t)} + A\sigma^2}{jj^{\prime}}};  \sum_{i,i^{\prime}, j, j^{\prime}} \rho_i\rho_{i^{\prime}}\lambda_j\lambda_{j^\prime} \frac{ii^{\prime}E^{(t)} + A\sigma^2}{jj^{\prime}}\Bigg);\\
v_{i\to a}^{(t+1)} &\approx h_{\text{var}}\Bigg(\c_i + z\sum_{i,i^{\prime}, j, j^{\prime}} \rho_i\rho_{i^{\prime}}\lambda_j\lambda_{j^\prime} \sqrt{\frac{ii^{\prime}E^{(t)} + A\sigma^2}{jj^{\prime}}};  \sum_{i,i^{\prime}, j, j^{\prime}}\rho_i\rho_{i^{\prime}}\lambda_j\lambda_{j^\prime} \frac{ii^{\prime}E^{(t)} + A\sigma^2}{jj^{\prime}}\Bigg).
\end{align}
Where $h_{\text{mean}}$ and $h_{\text{var}}$ are given by, 
\begin{align*}
h_{\text{mean}}(\mu; v) & = \lim_{\beta \to \infty} \frac{\int x_i e^{-\beta f(x_i)} e^{-\frac{\beta(x_i - \mu)^2}{2v}} dx_i}{\int e^{-\beta f(x_i)} e^{-\frac{\beta(x_i - \mu)^2}{2v}} dx_i}; \quad h_{\text{var}}(\mu;v) = \lim_{\beta \to \infty} \frac{\int x_i^2 e^{-\beta f(x_i)} e^{-\frac{\beta(x_i - \mu)^2}{2v}}dx_i}{\int e^{-\beta f(x_i)} e^{-\frac{\beta(x_i - \mu)^2}{2v}} dx_i} - h_{\text{mean}}(\mu;v)
\end{align*}
By plugging equations (21) and (22) in (8) and (9) yields the following,
\begin{align}
E^{(t+1)} &= \textbf{E}_{\text{prior}(s)}\textbf{E}_z\Bigg[h_{\text{mean}}\Bigg(s + z\sum_{i,i^{\prime}, j, j^{\prime}} \rho_i\rho_{i^{\prime}}\lambda_j\lambda_{j^\prime} \sqrt{\frac{ii^{\prime}E^{(t)} + A\sigma^2}{jj^{\prime}}};  \sum_{i,i^{\prime}, j, j^{\prime}} \rho_i\rho_{i^{\prime}}\lambda_j\lambda_{j^\prime} \frac{ii^{\prime}E^{(t)} + A\sigma^2}{jj^{\prime}}\Bigg) - s\Bigg]^2;\\
V^{(t+1)} &= \textbf{E}_{\text{prior}(s)}\textbf{E}_zh_{\text{var}}\Bigg(s + z\sum_{i,i^{\prime}, j, j^{\prime}} \rho_i\rho_{i^{\prime}}\lambda_j\lambda_{j^\prime} \sqrt{\frac{ii^{\prime}E^{(t)} + A\sigma^2}{jj^{\prime}}};  \sum_{i,i^{\prime}, j, j^{\prime}} \rho_i\rho_{i^{\prime}}\lambda_j\lambda_{j^\prime} \frac{ii^{\prime}E^{(t)} + A\sigma^2}{jj^{\prime}}\Bigg).
\end{align}
By setting $f(\bm{\c}) = \beta\norm{\bm{\c}}_1$, we enforce the returned solutions to be sparse. This is equivalent to choosing Laplacian prior for $\bm{\c}$. Following \cite{mpcs} in the noiseless case, equations (23) and (24) reduce to equations (10) and (11). 

\section{Relaxation of Message-passing convergence constraint}
\label{sec:mp-relax}
In this section we sketch the proof of Theorem 3.1, refer to \cite{hang_de} for more details of the proof. The derivation of necessary conditions for $\lim_{t\to\infty} (E^{(t)}, V^{(t)}) = (0, 0)$ can be split into two parts:
\begin{itemize}
\item \textbf{Part 1}. Showing that $(0,0)$ is a fixed point of the DE update equation.
\item \textbf{Part 2}. Necessary conditions for DE update equations to converge in the neighborhood of $(0,0)$. 
\end{itemize}
By substituting $(E^{(t)}, V^{(t)}) = (0, 0)$ we can see that it is indeed a fixed point. We begin part 2 by analyzing the functions $\delta_E^{(t)} = E^{(t+1)} - E^{(t)}$ and $\delta_V^{(t)} = V^{(t+1)} - V^{(t)}$. Let us define the functions $\Psi_E$ and $\Psi_V$ as follows,
\begin{align*}
\Psi_E(E^{(t)}; V^{(t)}) &= \textbf{E}_{\text{prior}(s)}\textbf{E}_{z\sim\mathcal{N}(0,1)}\bigg[\text{prox}\Big(s + a_1z\sqrt{E^{(t)}};\beta a_2V^{(t)}\Big) - s\bigg]^2;\\ 
\Psi_V(E^{(t)}; V^{(t)}) &= \textbf{E}_{\text{prior}(s)}\textbf{E}_{z\sim\mathcal{N}(0,1)}\bigg[\beta a_2V^{(t)}\text{prox}^{\prime}\Big(s + a_1z\sqrt{E^{(t)}}; \beta a_2V^{(t)}\Big)\bigg]^2.
\end{align*}
Taking the Taylor expansion of $\delta_E^{(t+1)}$ and $\delta_V^{(t+1)}$ and dropping the higher order terms we obtain,
$$
\begin{bmatrix}
\delta_E^{(t+1)}\\
\delta_V^{(t+1)}
\end{bmatrix}
=
\underbrace{\begin{bmatrix}
\Big(\frac{\partial \Psi_E(E, V)}{\partial E}\Big)^{(t)} & \Big(\frac{\partial \Psi_E(E, V)}{\partial V}\Big)^{(t)}\\
\Big(\frac{\partial \Psi_V(E, V)}{\partial E}\Big)^{(t)} & \Big(\frac{\partial \Psi_V(E, V)}{\partial V}\Big)^{(t)}
\end{bmatrix}}_{=: \bm{L}^{(t)}}
\begin{bmatrix}
\delta_E^{(t)}\\
\delta_V^{(t)}
\end{bmatrix}
$$ 
For $\Psi_E$ and $\Psi_V$ to converge to $0$, we would want the operator norm of $\bm{L}^{(t)}$ to be less than 1, i.e., $\inf_t \norm{\bm{L}^{(t)}} \leq 1$. Since 
$$\norm{\bm{L}^{(t)}} = \max
\Bigg[ \Big(\frac{\partial \Psi_E(E, V)}{\partial E}\Big)^{(t)}, \Big(\frac{\partial \Psi_V(E, V)}{\partial V}\Big)^{(t)}\Bigg]. $$
We can restrict the lower bounds of the individual terms to be less than 1. This would result in
$$a_1^2 \leq \frac{p^2}{k^2}, \quad a_2 \leq \frac{p^2}{k^2\beta}.$$

\section{Relaxation of Constraints for Preferential Sensing}

In this section, we provide details for the relaxation of requirements (2) and (3) for preferential sensing. In this regime, we separately track the average error and the variance of the HH, HL (LH), and LL parts of the covariance matrix separately. The quantities $E_{HH}^{(t)}, V_{HH}^{(t)}, E_{HL}^{(t)}, V_{HL}^{(t)}$, and $E_{LL}^{(t)}, V_{LL}^{(t)}$ are defined as described in section \ref{ssec:de-unequal} and following the procedure described in appendix \ref{sec:DE-update} yields equation \ref{eq;pref-req-param}. Let us now define the following quantities
\begin{align*}
E_{HH}^{(t+1)} &= \textbf{E}_{\text{prior}(s)}\textbf{E}_{z\sim\mathcal{N}(0,1)}\bigg[\text{prox}\Big(s + zb_{HH, 1}^{(t)}; b_{HH, 2}^{(t)}\Big) - s\bigg]^2\\ 
    &\triangleq \Psi_{E,HH}\Big(E_{HH}^{(t)}, V_{HH}^{(t)}, E_{HL}^{(t)}, V_{HL}^{(t)},E_{LL}^{(t)}, V_{LL}^{(t)}\Big); \\
E_{HL}^{(t+1)} &= \textbf{E}_{\text{prior}(s)}\textbf{E}_{z\sim\mathcal{N}(0,1)}\bigg[\text{prox}\Big(s + zb_{HL, 1}^{(t)}; b_{HL, 2}^{(t)}\Big) - s\bigg]^2\\ 
    &\triangleq \Psi_{E,HL}\Big(E_{HH}^{(t)}, V_{HH}^{(t)}, E_{HL}^{(t)}, V_{HL}^{(t)},E_{LL}^{(t)}, V_{LL}^{(t)}\Big); \\
E_{LL}^{(t+1)} &= \textbf{E}_{\text{prior}(s)}\textbf{E}_{z\sim\mathcal{N}(0,1)}\bigg[\text{prox}\Big(s + zb_{LL, 1}^{(t)}; b_{LL, 2}^{(t)}\Big) - s\bigg]^2\\ 
    &\triangleq \Psi_{E,LL}\Big(E_{HH}^{(t)}, V_{HH}^{(t)}, E_{HL}^{(t)}, V_{HL}^{(t)},E_{LL}^{(t)}, V_{LL}^{(t)}\Big);
\end{align*}
Similarly,
\begin{align*}
V_{HH}^{(t+1)} &= \textbf{E}_{\text{prior}(s)}\textbf{E}_{z\sim\mathcal{N}(0,1)}\bigg[b_{HH, 2}^{(t)}\text{prox}^{\prime}\Big(s + zb_{HH, 1}^{(t)};  b_{HH, 2}^{(t)}\Big)\bigg]\\
    &\triangleq \Psi_{V,HH}\Big(E_{HH}^{(t)}, V_{HH}^{(t)}, E_{HL}^{(t)}, V_{HL}^{(t)},E_{LL}^{(t)}, V_{LL}^{(t)}\Big); \\
V_{HL}^{(t+1)} &= \textbf{E}_{\text{prior}(s)}\textbf{E}_{z\sim\mathcal{N}(0,1)}\bigg[b_{HH, 2}^{(t)}\text{prox}^{\prime}\Big(s + zb_{HL, 1}^{(t)};  b_{HL, 2}^{(t)}\Big)\bigg]\\
    &\triangleq \Psi_{V,HL}\Big(E_{HH}^{(t)}, V_{HH}^{(t)}, E_{HL}^{(t)}, V_{HL}^{(t)},E_{LL}^{(t)}, V_{LL}^{(t)}\Big); \\
V_{LL}^{(t+1)} &= \textbf{E}_{\text{prior}(s)}\textbf{E}_{z\sim\mathcal{N}(0,1)}\bigg[b_{HH, 2}^{(t)}\text{prox}^{\prime}\Big(s + zb_{LL, 1}^{(t)};  b_{LL, 2}^{(t)}\Big)\bigg]\\
    &\triangleq \Psi_{V,HL}\Big(E_{HH}^{(t)}, V_{HH}^{(t)}, E_{HL}^{(t)}, V_{HL}^{(t)},E_{LL}^{(t)}, V_{LL}^{(t)}\Big); \\
\end{align*}
We now define $\delta_{E, HH}^{(t)}, \delta_{E, HL}^{(t)}, \delta_{E, LL}^{(t)}$, and $\delta_{V, HH}^{(t)}, \delta_{V, HL}^{(t)}, \delta_{V, LL}^{(t)}$ in a similar manner to that in appendix \ref{sec:mp-relax}. 

\subsection{Relaxation of Requirement 2}
We use the shorthand, $\Psi_{V, HH}^{(t)} = \triangleq \Psi_{V,HL}\Big(E_{HH}^{(t)}, V_{HH}^{(t)}, E_{HL}^{(t)}, V_{HL}^{(t)},E_{LL}^{(t)}, V_{LL}^{(t)}\Big)$ for ease of notation.  Approximate $\delta_{V, HH}^{(t)}$ using its First-order Taylor series expansion, we get
\begin{align*}
    \delta_{V, HH}^{(t+1)} &= \Psi_{V,HH}^{(t+1)} - \Psi_{V,HH}^{(t)}\\
        &= \Bigg(\frac{\partial \Psi_{V, HH}(\cdot)}{\partial E_{HH}}\Bigg)^{(t)} \delta_{E, HH}^{(t)} + \Bigg(\frac{\partial \Psi_{V, HH}(\cdot)}{\partial E_{HL}}\Bigg)^{(t)} \delta_{E, HL}^{(t)} + \Bigg(\frac{\partial \Psi_{V, HH}(\cdot)}{\partial E_{LL}}\Bigg)^{(t)} \delta_{E, LL}^{(t)}\\
        & \quad + \Bigg(\frac{\partial \Psi_{V, HH}(\cdot)}{\partial V_{HH}}\Bigg)^{(t)} \delta_{V, HH}^{(t)} + \Bigg(\frac{\partial \Psi_{V, HH}(\cdot)}{\partial V_{HL}}\Bigg)^{(t)} \delta_{V, HL}^{(t)} + \Bigg(\frac{\partial \Psi_{V, HH}(\cdot)}{\partial V_{LL}}\Bigg)^{(t)} \delta_{V, LL}^{(t)}\\
        & \quad + O\bigg(\Big(\delta_{V, HH}^{(t)}\Big)^2\bigg) + O\bigg(\Big(\delta_{V, HL}^{(t)}\Big)^2\bigg) + O\bigg(\Big(\delta_{V, LL}^{(t)}\Big)^2\bigg)
\end{align*}
Following the same template as appendix \ref{sec:mp-relax}, the derivation consists of two parts:
\begin{enumerate}[label=Part \Roman*]
    \item Verify that $(0,0,0)$ is a fixed point. Which is a trivial task. 
    \item Show that the DE equations w.r.t to $V_{HH}^{(t)}, V_{HL}^{(t)}, V_{LL}^{(t)}$ converges within a proximity of the origin. 
\end{enumerate}
It can be trivially checked that part I is true. We now focus our attention to part II. Consider the region where $V_{HH}^{(t)}, V_{HL}^{(t)}, V_{LL}^{(t)}$, in this case, we can ignore the quadratic terms in the above equation. By exploiting the fact that $\partial\Psi_{V, HH}/\partial E_{HH} = \partial\Psi_{V, HH}/\partial E_{HL} = \partial\Psi_{V, HH}/\partial E_{LL} = 0$, we obtain the following. 
\[
\begin{bmatrix}
    \delta_{V, HH}^{(t+1)}\\
    \delta_{V, HL}^{(t+1)}\\
    \delta_{V, LL}^{(t+1)}
\end{bmatrix} 
=
\underbrace{\begin{bmatrix}
    \Big(\frac{\partial\Psi_{V, HH}}{\partial V_{HH}}\Big)^{(t)} & \Big(\frac{\partial\Psi_{V, HH}}{\partial V_{HL}}\Big)^{(t)} & \Big(\frac{\partial\Psi_{V, HH}}{\partial V_{LL}}\Big)^{(t)}\\
    \Big(\frac{\partial\Psi_{V, HL}}{\partial V_{HH}}\Big)^{(t)} & \Big(\frac{\partial\Psi_{V, HL}}{\partial V_{HL}}\Big)^{(t)} & \Big(\frac{\partial\Psi_{V, HL}}{\partial V_{LL}}\Big)^{(t)}\\
    \Big(\frac{\partial\Psi_{V, LL}}{\partial V_{HH}}\Big)^{(t)} & \Big(\frac{\partial\Psi_{V, LL}}{\partial V_{HL}}\Big)^{(t)} & \Big(\frac{\partial\Psi_{V, LL}}{\partial V_{LL}}\Big)^{(t)}
\end{bmatrix}}_{\bm{L}_V^{(t)}}
\begin{bmatrix}
    \delta_{V, HH}^{(t)}\\
    \delta_{V, HL}^{(t)}\\
    \delta_{V, LL}^{(t)}
\end{bmatrix} 
\]
To make the LHS convergent we require $\inf_t \norm{\bm{L}_V^{(t)}}_{OP} \leq 1$. We now lower each term in the first row of $\bm{L}_V^{(t)}$ similar to what was done in appendix \ref{sec:mp-relax}, hence we omit the details. We then obtain,
\begin{align*}
    \Big(\frac{\partial\Psi_{V, HH}}{\partial V_{HH}}\Big)^{(t)} & \geq \frac{k_{HH}\beta_{HH}}{n_{HH}} \bigg(\sum_{\ell}\frac{\lambda_{H,\ell}}{\ell}\bigg)^2 \bigg(\sum_{i}i\rho_{H,i}\bigg)^2\\
    \Big(\frac{\partial\Psi_{V, HH}}{\partial V_{HL}}\Big)^{(t)} & \geq \frac{k_{HH}\beta_{HH}}{n_{HH}} \bigg(\sum_{\ell}\frac{\lambda_{H,\ell}}{\ell}\bigg)^2 \bigg(\sum_{i}i\rho_{H,i}\bigg)\bigg(\sum_{j}j\rho_{L,j}\bigg)\\
    \Big(\frac{\partial\Psi_{V, HH}}{\partial V_{LL}}\Big)^{(t)} & \geq \frac{k_{HH}\beta_{HH}}{n_{HH}} \bigg(\sum_{\ell}\frac{\lambda_{H,\ell}}{\ell}\bigg)^2 \bigg(\sum_{i}i\rho_{L,i}\bigg)^2\\
\end{align*}
Following the same procedure for the second row, we get
\begin{align*}
    \Big(\frac{\partial\Psi_{V, HL}}{\partial V_{HH}}\Big)^{(t)} & \geq \frac{k_{HL}\beta_{HL}}{n_{HL}} \bigg(\sum_{\ell}\frac{\lambda_{H,\ell}}{\ell}\bigg)\bigg(\sum_{k}\frac{\lambda_{L,k}}{k}\bigg) \bigg(\sum_{i}i\rho_{H,i}\bigg)^2\\
    \Big(\frac{\partial\Psi_{V, HL}}{\partial V_{HL}}\Big)^{(t)} & \geq \frac{k_{HL}\beta_{HL}}{n_{HL}} \bigg(\sum_{\ell}\frac{\lambda_{H,\ell}}{\ell}\bigg)\bigg(\sum_{k}\frac{\lambda_{L,k}}{k}\bigg)\bigg(\sum_{i}i\rho_{H,i}\bigg)\bigg(\sum_{j}j\rho_{L,j}\bigg)\\
    \Big(\frac{\partial\Psi_{V, HL}}{\partial V_{LL}}\Big)^{(t)} & \geq \frac{k_{HL}\beta_{HL}}{n_{HL}} \bigg(\sum_{\ell}\frac{\lambda_{H,\ell}}{\ell}\bigg)\bigg(\sum_{k}\frac{\lambda_{L,k}}{k}\bigg) \bigg(\sum_{i}i\rho_{L,i}\bigg)^2\\
\end{align*}
And finally for row 3 we get,
\begin{align*}
    \Big(\frac{\partial\Psi_{V, LL}}{\partial V_{HH}}\Big)^{(t)} & \geq \frac{k_{LL}\beta_{LL}}{n_{LL}} \bigg(\sum_{\ell}\frac{\lambda_{L,\ell}}{\ell}\bigg)^2 \bigg(\sum_{i}i\rho_{H,i}\bigg)^2\\
    \Big(\frac{\partial\Psi_{V, LL}}{\partial V_{HL}}\Big)^{(t)} & \geq \frac{k_{LL}\beta_{LL}}{n_{LL}} \bigg(\sum_{\ell}\frac{\lambda_{L,\ell}}{\ell}\bigg)^2 \bigg(\sum_{i}i\rho_{H,i}\bigg)\bigg(\sum_{j}j\rho_{L,j}\bigg)\\
    \Big(\frac{\partial\Psi_{V, LL}}{\partial V_{LL}}\Big)^{(t)} & \geq \frac{k_{LL}\beta_{LL}}{n_{LL}} \bigg(\sum_{\ell}\frac{\lambda_{L,\ell}}{\ell}\bigg)^2 \bigg(\sum_{i}i\rho_{L,i}\bigg)^2\\
\end{align*}
Equation (\ref{eq:req2}) is then obtained by enforcing the condition on the operator norm on the above inequalities. 

\subsection{Relaxation of Requirement 3}
The basic idea remains the same as in the previous subsection. We linearize the DE update equation with Taylor expansion and enforce the difference $\delta_{E, HH}^{(t)}$ to decrease faster than $\delta_{E, HL}^{(t)}$ and $\delta_{E, LL}^{(t)}$. That is,
\begin{align}
    \Big(\frac{\partial\Psi_{E, HH}}{\partial E_{HH}}\Big)^{(t)} & \leq \Big(\frac{\partial\Psi_{E, HL}}{\partial E_{HH}}\Big)^{(t)};\\
    \Big(\frac{\partial\Psi_{E, HH}}{\partial E_{HL}}\Big)^{(t)} & \leq \Big(\frac{\partial\Psi_{E, HL}}{\partial E_{HL}}\Big)^{(t)};\\
    \Big(\frac{\partial\Psi_{E, HH}}{\partial E_{LL}}\Big)^{(t)} & \leq \Big(\frac{\partial\Psi_{E, HL}}{\partial E_{LL}}\Big)^{(t)}.
\end{align}
And,
\begin{align}
    \Big(\frac{\partial\Psi_{E, HH}}{\partial E_{HH}}\Big)^{(t)} & \leq \Big(\frac{\partial\Psi_{E, LL}}{\partial E_{HH}}\Big)^{(t)};\\
    \Big(\frac{\partial\Psi_{E, HH}}{\partial E_{HL}}\Big)^{(t)} & \leq \Big(\frac{\partial\Psi_{E, LL}}{\partial E_{HL}}\Big)^{(t)};\\
    \Big(\frac{\partial\Psi_{E, HH}}{\partial E_{LL}}\Big)^{(t)} & \leq \Big(\frac{\partial\Psi_{E, LL}}{\partial E_{LL}}\Big)^{(t)}.
\end{align}
Following the same logic as the previous subsection, we can lower-bound each of the gradients in the above inequalities. We then obtain,
\begin{align*}
    \Big(\frac{\partial\Psi_{E, HH}}{\partial E_{HH}}\Big)^{(t)} & \geq \frac{k_{HH}}{n_{HH}} \bigg(\sum_{\ell}\frac{\lambda_{H,\ell}}{\sqrt{\ell}}\bigg)^4 \bigg(\sum_{i}\sqrt{i}\rho_{H,i}\bigg)^4\\
    \Big(\frac{\partial\Psi_{E, HH}}{\partial E_{HL}}\Big)^{(t)} & \geq \frac{k_{HH}}{n_{HH}} \bigg(\sum_{\ell}\frac{\lambda_{H,\ell}}{\sqrt{\ell}}\bigg)^4 \bigg(\sum_{i}\sqrt{i}\rho_{H,i}\bigg)^2\bigg(\sum_{j}\sqrt{j}\rho_{L,j}\bigg)^2\\
    \Big(\frac{\partial\Psi_{E, HH}}{\partial E_{LL}}\Big)^{(t)} & \geq \frac{k_{HH}}{n_{HH}} \bigg(\sum_{\ell}\frac{\lambda_{H,\ell}}{\sqrt{\ell}}\bigg)^4 \bigg(\sum_{i}\sqrt{i}\rho_{L,i}\bigg)^4\\
\end{align*}
And,
\begin{align*}
    \Big(\frac{\partial\Psi_{E, HL}}{\partial E_{HH}}\Big)^{(t)} & \geq \frac{k_{HL}}{n_{HL}} \bigg(\sum_{\ell}\frac{\lambda_{H,\ell}}{\sqrt{\ell}}\bigg)^2\bigg(\sum_{\ell}\frac{\lambda_{L,\ell}}{\sqrt{\ell}}\bigg)^2 \bigg(\sum_{i}\sqrt{i}\rho_{H,i}\bigg)^4\\
    \Big(\frac{\partial\Psi_{E, HL}}{\partial E_{HL}}\Big)^{(t)} & \geq \frac{k_{HL}}{n_{HL}} \bigg(\sum_{\ell}\frac{\lambda_{H,\ell}}{\sqrt{\ell}}\bigg)^2\bigg(\sum_{\ell}\frac{\lambda_{L,\ell}}{\sqrt{\ell}}\bigg)^2 \bigg(\sum_{i}\sqrt{i}\rho_{H,i}\bigg)^2\bigg(\sum_{j}\sqrt{j}\rho_{L,j}\bigg)^2\\
    \Big(\frac{\partial\Psi_{E, HL}}{\partial E_{LL}}\Big)^{(t)} & \geq \frac{k_{HL}}{n_{HL}} \bigg(\sum_{\ell}\frac{\lambda_{H,\ell}}{\sqrt{\ell}}\bigg)^2 \bigg(\sum_{\ell}\frac{\lambda_{L,\ell}}{\sqrt{\ell}}\bigg)^2\bigg(\sum_{i}\sqrt{i}\rho_{L,i}\bigg)^4\\
\end{align*}
Finally, 
\begin{align*}
    \Big(\frac{\partial\Psi_{E, LL}}{\partial E_{HH}}\Big)^{(t)} & \geq \frac{k_{LL}}{n_{LL}} \bigg(\sum_{\ell}\frac{\lambda_{L,\ell}}{\sqrt{\ell}}\bigg)^4 \bigg(\sum_{i}\sqrt{i}\rho_{H,i}\bigg)^4\\
    \Big(\frac{\partial\Psi_{E, LL}}{\partial E_{HL}}\Big)^{(t)} & \geq \frac{k_{LL}}{n_{LL}} \bigg(\sum_{\ell}\frac{\lambda_{L,\ell}}{\sqrt{\ell}}\bigg)^4 \bigg(\sum_{i}\sqrt{i}\rho_{H,i}\bigg)^2\bigg(\sum_{j}\sqrt{j}\rho_{L,j}\bigg)^2\\
    \Big(\frac{\partial\Psi_{E, LL}}{\partial E_{LL}}\Big)^{(t)} & \geq \frac{k_{LL}}{n_{LL}} \bigg(\sum_{\ell}\frac{\lambda_{L,\ell}}{\sqrt{\ell}}\bigg)^4 \bigg(\sum_{i}\sqrt{i}\rho_{L,i}\bigg)^4\\
\end{align*}
Combining this with inequalities (38)-(43) yields inequality (\ref{eq:req3-2}). 

\section{Graph Structure Recovery (Regular sensing)}
\begin{figure}[t]
\centering
\includegraphics[width=0.8\linewidth]{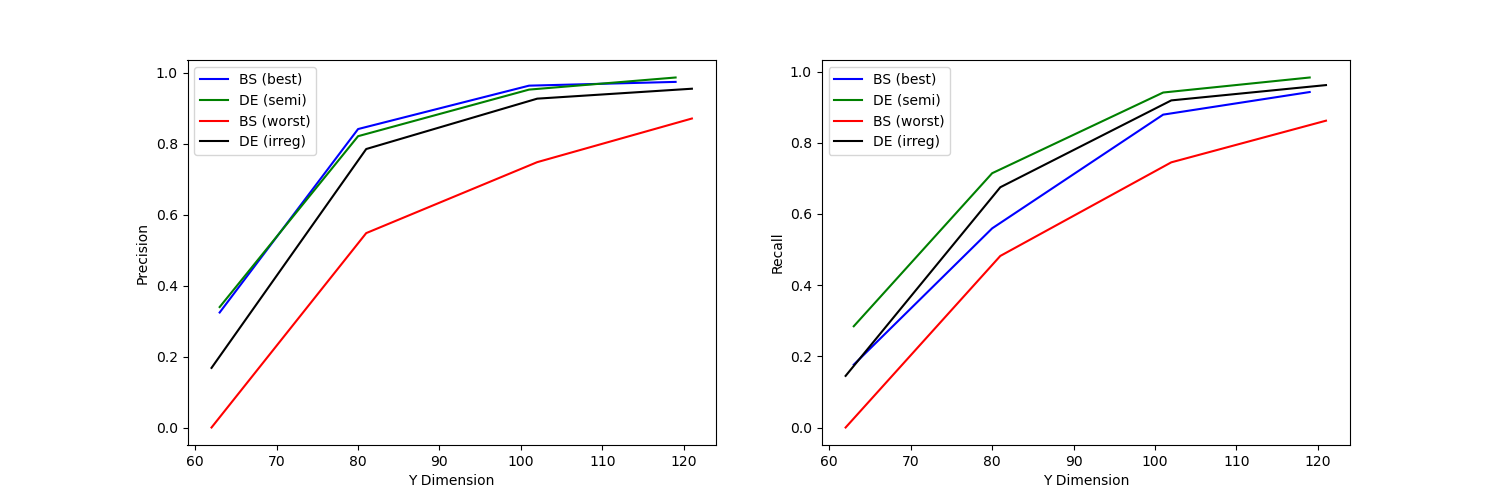}
\caption{Comparison of the performance of the proposed sensing system with that of \cite{mksetch} (BS - best, BS - worst lines in the plots). Number of nodes, $p = 200$. The performance is evaluated with respect to \textit{precision} and \textit{recall} of the edges in the graph.}
\label{fig:graph_comp_200}
\end{figure}

Here we compare the performance of the proposed regular sensing matrix on graph structure recovery task with the sensing system proposed by \citet{mksetch}. The sensing systems are evaluated with respect to: (i) MAE, (ii) Precision, and (iii) Recall. We can see from Figure \ref{fig:graph_comp_200} that the relative performance between the two systems is similar to the behavior exhibited on the covariance recovery task. That is, the two sensing systems are at an equal footing when the baseline is tuned. 


\end{document}